# Modèle multi-échelles pour les services dans les VANETs

Marie-Ange Lebre[#*], Frédéric Le Mouël[#], Eric Ménard[*]

[#]*Laboratoire CITI, INSA de Lyon, Villeurbanne, France*
[*]*Valeo, Etudes Electroniques, Créteil, France*
marie-ange.lebre@insa-lyon.fr

*Abstract*— Le déploiement et l'utilisation de services dans les VANETs sont en pleine expansion. Actuellement ils sont principalement liés à la sécurité du conducteur. Dans le futur, le déploiement des applications permettra de considérer la voiture comme le prolongement de l'utilisateur, comme le smartphone actuellement. Nous proposons un modèle multi-échelles relevant ce défi d'intégration VANET / application / utilisateur.

## I. Introduction

Trouver des applications pertinentes au sein des 'Vehicular ad hoc Networks' (VANETs) est un enjeu majeur de la popularité et de l'utilisabilité d'un véhicule auprès des consommateurs. Le dynamisme du réseau, les contraintes de délai des services et le rôle central de l'utilisateur représentent les clés de la voiture communicante de demain.

L'aspect original de ce travail est de modéliser et d'analyser la dynamique des inter-connectivités des véhicules, la dynamique des intéractions sociales des utilisateurs et la dynamique du déploiement d'application en utilisant la théorie des graphes. Cet article dresse un bref état de l'art sur les applications pensées pour les VANETs, puis détaille notre modèle multi-échelles, et enfin conclue et donne des perspectives futures.

## II. Contexte de l'etude

Nous avons effectué une veille technologique autour du thème 'smart car' et des projets V2X (Vehicle to X communication [1], [2]) allant de l'assistance à la conduite jusqu'à la future voiture sociale et communicante.

Nous avons défini un classement de tous ces projets en fonction de la complexité des applications autour du véhicule et de l'utilisateur. Ceci a mis en évidence un vide entre l'utilisateur et la voiture. En effet, certains projets sont centrés exclusivement sur le véhicule autour du thème de la sécurité et d'autres sont centrés sur l'utilisateur avec l'intégration du web et des applications mobiles dans l'habitacle. La vue des services est soit centrée véhicule, soit centrée utilisateur, il n'y a pas réellement de lien applicatif entre le réseau véhiculaire et le réseau d'utilisateur.

## III. Le modèle multi-echelles

Un modèle multi-échelles apparaît alors, composé de trois graphes : véhiculaire, applicatif et utilisateur.

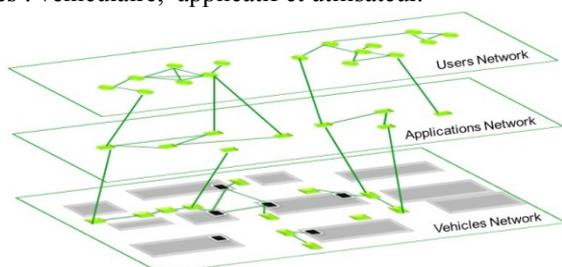

Figure 1. Modèle Multi-échelles.

Le graphe d'application est le plus délicat à modéliser car il est vaste et difficile à cerner de part son hétérogénéité. La congestion étant actuellement un réel problème, nous avons donc décidé de nous concentrer sur des applications favorisant la fluidité du trafic. Nous avons choisi 'Green Wave' pour optimiser la fluidité le long d'un chemin selon différents critères choisis par l'utilisateur et la fluidité aux intersections.

Les trois graphes étant très différents, la nomenclature des liens inter-niveaux se révèle être complexe. Un état de l'art sur la modélisation des réseaux sociaux avec la théorie des graphes nous a permis d'identifier différents types de lien. Les liens explicites qui ne possèdent aucune ambiguïté dans leur description ; les liens implicites induits par des circonstances particulières et les liens associatifs qui représentent une association de nœuds ayant des intérêts communs dans un contexte donné (voir la figure 2).

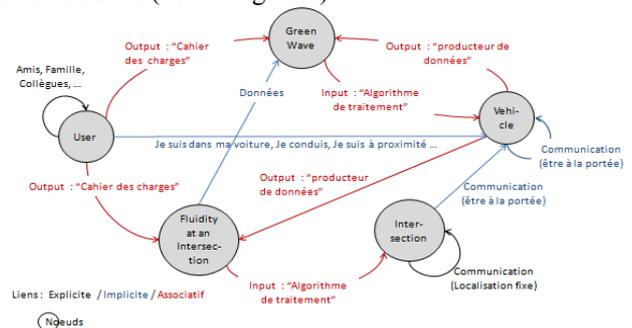

Figure 2. Nomenclature du modèle

Notre travail consistera à analyser ces graphes de manière indépendante pour extraire et découvrir des propriétés et d'en garantir sur le modèle multi-niveaux. Pour cela nous mettrons en place un algorithme autonome et distribué dans le réseau véhiculaire grâce à la communication V2X via nos deux applications sur la fluidité. Ce dernier va nous permettre d'assurer des propriétés locales à chaque niveau et grâce à la dissémination d'information et à la dynamique du réseau nous pourrons extraire et garantir des propriétés globales.

## IV. Conclusions et perspectives

Dans ce bref article nous présentons un modèle permettant de considérer le réseau véhiculaire, le réseau d'applications et le réseau d'utilisateurs dans son ensemble grâce à la théorie des graphes. La prochaine étape sera les simulations et la validation des propriétés observées grâce à des tests réels.